\documentclass[prl,superscriptaddress,floatfix,amsmath,footinbib,amssymb,twocolumn]{revtex4}
\usepackage{amssymb}
\usepackage{amsmath}
\usepackage{amsfonts}
\usepackage{tikz}
\usepackage{bm}

\usepackage{epsfig}
\usepackage{t1enc}
\usepackage{soul}
\usepackage{color}

\usepackage{mathrsfs}
\usepackage{url}
\usepackage[all]{xy}
\usetikzlibrary{calc,patterns,angles,quotes}

\newcommand*\diff{\mathop{}\!\mathrm{d}}

\begin{document}

\title{Three-line derivation of the Thomas precession}

\author{Pawe\l  ~Lewulis\thanks{Supported by NCN Preludium 11, 2016/21/N/ST1/02599.}}\affiliation{Institute of Mathematics, Polish Academy of Sciences, \'Sniadeckich 8, 00-656, Warsaw, Poland}

\author{Andrzej Dragan}
\affiliation{Institute of Theoretical Physics, University of Warsaw, Pasteura 5, 02-093 Warsaw, Poland}
\affiliation{Centre for Quantum Technologies, National University of Singapore, 3 Science Drive 2, 117543 Singapore, Singapore}
\date{\today}
\maketitle

Recently \cite{halfpage}, a half-page derivation of the Thomas precession effect has been presented (see also all the references therein for the history and prehistory of the problem). In this short note we provide a much simpler, three-line derivation of that effect which, to our knowledge, is the simplest in the literature. 

\begin{figure}
   \begin{tikzpicture}
	\foreach \Point/\PointLabel in {  (2.125,0.5)/B, (2.5,2)/B', (0.625,0.5)/A  }
	\draw[fill=black] \Point circle (0.1) node[below=0.35cm, left=-0.1cm] {$\PointLabel$};
		\coordinate (pivot) at (2,2);
     \draw [ultra thick] [->] (2.125,0.5) -- (3.125,0.5) node[right]{$\bm{v}$};
    	\draw [ultra thick] [->] (2.5,2) -- (3.5,2.8) node [above=0.25cm ,right=-0.15cm]{$\bm{v} + \diff  \bm{v}$} ;
   	\draw [ultra thick] [->] (2.5,2) -- (2.78,3.125) node  [above=0.3cm ,right=-0.3cm]{$\diff \bm{v}'$};
    \end{tikzpicture}
  \caption{\label{schemat}Relative velocities between Alice's frame $A$, Bob's frame $B$ and new Bob's frame $B'$.}
\end{figure}
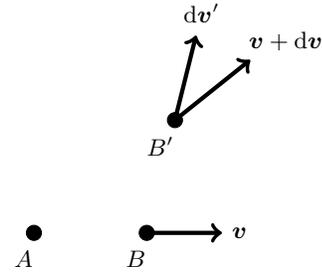

Consider Bob $B$ moving relative to Alice's frame $A$ with a relativistic velocity $\boldsymbol{v}$, as shown in Fig.~\ref{schemat}. If Bob's velocity changes by an infinitesimal value $\text{d}\boldsymbol{v}'$ relative to his initial reference frame $B$, Alice will observe his new velocity to be some $\boldsymbol{v}+\text{d}\boldsymbol{v}$. The new Bob's frame $B'$ is now rotated relative to Alice's frame by a Thomas--Wigner angle $\text{d}\boldsymbol{\Omega}$. In the non-relativistic theory we have $\text{d}\boldsymbol{v}' = \text{d}\boldsymbol{v}$ and the rotation is absent. Therefore, the Thomas--Wigner angle can be interpreted as an angle between the relativistic velocity of $B'$ with respect to $A$, $\boldsymbol{v}+\text{d}\boldsymbol{v}$, and it's non-relativistic approximation, $\boldsymbol{v}+\text{d}\boldsymbol{v}'$:
\begin{equation}
\label{angle}
\text{d}\boldsymbol{\Omega} = \frac{(\boldsymbol{v} + \text{d}\boldsymbol{v'})\times(\boldsymbol{v} + \text{d}\boldsymbol{v})}{v^2} = \frac{1}{v^2} (\boldsymbol{v}\times\text{d}\boldsymbol{v} - \boldsymbol{v}\times\text{d}\boldsymbol{v}'). 
\end{equation}

Let us use a relativistic transformation of the perpendicular velocity component, $\boldsymbol{u}'_\perp = \frac{\boldsymbol{u}_\perp \sqrt{1-v^2}}{1-\boldsymbol{u}\cdot\boldsymbol{v}}$. We will consider a velocity transformation from $A$ to $B$. Here we substitute $\boldsymbol{u} \rightarrow \boldsymbol{v} + \text{d}\boldsymbol{v}$ and $\boldsymbol{u}' \rightarrow \text{d}\boldsymbol{v}'$. Next, we take the cross product of the resulting formula with $\boldsymbol{v}$:

\begin{equation}
\label{prod}
\boldsymbol{v}\times\text{d}\boldsymbol{v}' = \boldsymbol{v}\times \frac{(\boldsymbol{v} + \text{d}\boldsymbol{v}) \sqrt{1-v^2}}{1-(\boldsymbol{v} + \text{d}\boldsymbol{v})\cdot\boldsymbol{v}} = \frac{\boldsymbol{v}\times\text{d}\boldsymbol{v}}{\sqrt{1-v^2}},
\end{equation}
where we used the fact that the parallel (to $\boldsymbol{v}$) components of velocities do not contribute to the vector product with $\boldsymbol{v}$, and neglected higher order corrections from the denominator. Substituting Eq.~\eqref{prod} into Eq.~\eqref{angle} and dividing by an infinitesimal time $\text{d}t$ we obtain the Thomas precession rate:

\begin{equation}
\dot{\boldsymbol{\Omega}} = -\frac{1}{v^2}\left(\frac{1}{\sqrt{1-v^2}} -1 \right) \boldsymbol{v}\times\dot{\boldsymbol{v}}.
\end{equation}

\end{document}